\documentclass{article}

\usepackage{arxiv}

\usepackage[utf8]{inputenc} 
\usepackage[T1]{fontenc}    
\usepackage{hyperref}       
\usepackage{url}            
\usepackage{booktabs}       
\usepackage{amsfonts}       
\usepackage{nicefrac}       
\usepackage{microtype}      
\usepackage{cleveref}       
\usepackage{graphicx}
\usepackage[numbers]{natbib}
\usepackage{doi}
\usepackage{multirow}
\usepackage{array} 
\usepackage{placeins} 

\title{SPIDER: A Comprehensive Multi-Organ Supervised Pathology Dataset and Baseline Models}
\date{}

\usepackage{authblk}

\author[1]{Dmitry Nechaev\thanks{\texttt{dmitry@hist.ai}}}
\author[1]{Alexey Pchelnikov\thanks{\texttt{alex@hist.ai}}}
\author[1]{Ekaterina Ivanova\thanks{\texttt{kate@hist.ai}}}
\affil[1]{HistAI}

\renewcommand{\headeright}{SPIDER: A Supervised Pathology Dataset and Baseline Models}
\renewcommand{\undertitle}{}
\renewcommand{\shorttitle}{}

\begin{document}
\maketitle

\begin{abstract}
Advancing AI in computational pathology requires large, high-quality, and diverse datasets, yet existing public datasets are often limited in organ diversity, class coverage, or annotation quality. To bridge this gap, we introduce SPIDER (Supervised Pathology Image-DEscription Repository), the largest publicly available patch-level dataset covering multiple organ types, including Skin, Colorectal, Thorax, and Breast with comprehensive class coverage for each organ. SPIDER provides high-quality annotations verified by expert pathologists and includes surrounding context patches, which enhance classification performance by providing spatial context.

Alongside the dataset, we present baseline models trained on SPIDER using the Hibou-L foundation model as a feature extractor combined with an attention-based classification head. The models achieve state-of-the-art performance across multiple tissue categories and serve as strong benchmarks for future digital pathology research. Beyond patch classification, the model enables rapid identification of significant areas, quantitative tissue metrics, and establishes a foundation for multimodal approaches.

Both the dataset and trained models are publicly available to advance research, reproducibility, and AI-driven pathology development. Access them at:
\href{https://github.com/HistAI/SPIDER}{https://github.com/HistAI/SPIDER}.

\end{abstract}

\section{Introduction}

Foundation models such as Vision Transformers~\citep{dosovitskiy2021imageworth16x16words} (ViTs) have revolutionized computer vision by enabling highly effective transfer learning. Unlike traditional end-to-end training, these models leverage a two-stage pipeline where extensive pretraining on large-scale datasets extracts universal features, and task-specific fine-tuning adapts the model efficiently to new domains \citep{Tomczyk2024Machine}. This approach significantly reduces computational costs, making state-of-the-art models accessible for real-world applications.

Similarly, computational pathology (CPath) has increasingly embraced foundation vision models~\citep{Xu2023Vision}, leveraging their ability to efficiently generalize across diverse histopathological tasks. By fine-tuning these models on domain-specific datasets, researchers have achieved significant advancements in cancer subtyping, biomarker discovery, and automated diagnostics \citep{chanda2024neweracomputationalpathology}. 

Motivated by these advancements, we sought to explore whether foundation models could also be effectively applied to whole slide image (WSI) analysis, specifically for whole slide image segmentation. One widely explored approach is using unsupervised clustering on the features extracted by these models, with the goal of automatically grouping similar tissue structures without requiring explicit labels \citep{gildenblat2024segmentationfactorizationunsupervisedsemantic}. The idea is that foundation models, pretrained on vast datasets, can produce meaningful feature representations that naturally separate different tissue types.

In theory, this approach should provide a way to segment WSIs without extensive manual annotation, making it an attractive option for large-scale pathology applications. However, our internal experiments revealed significant limitations in practice.

First, while clustering does create distinct groups, their biological meaning is unclear without expert pathologists manually annotating them. Second, often times multiple clusters may represent the same morphology, leading to redundant groupings. This means additional post-processing is needed to merge similar clusters. But most importantly, the quality of these clusters is often inadequate for real-world diagnostic applications. While simple structures like fat are relatively easy to detect, complex and clinically significant features, such as high-grade adenocarcinoma, are poorly segmented. This limits the usefulness of unsupervised clustering in pathology workflows, especially when precision is critical.

Instead, a more effective approach is to use foundation models in a fully supervised manner. By creating a patch-level dataset with expert-annotated labels, we can fine-tune a pretrained foundation model without needing an extremely large dataset. Unlike training a model from scratch, which requires vast amounts of data and computational power, this method takes advantage of the model’s existing ability to generalize well. Even with a moderately sized dataset, the model can achieve strong performance due to the powerful feature representations learned during pretraining.

Building on this approach:
\begin{itemize}
    \item We introduce \textbf{SPIDER}, 
    \textbf{S}upervised \textbf{P}athology \textbf{I}mage-\textbf{DE}scription \textbf{R}epository, 
    the largest publicly available patch-level dataset for multiple organ types, including Skin, Colorectal, Thorax, and Breast with comprehensive class coverage for each organ.
    Annotations were generated using a semi-automatic pipeline with every patch being further validated by professional pathologists to ensure accuracy. Additionally, each patch includes a context field of view, offering expanded context to enhance classification performance.
    
    \item We present a \textbf{baseline model architecture} for patch-level classification, designed around the Hibou-L \citep{nechaev2024hiboufamilyfoundationalvision} foundation model. This architecture is capable of both patch classification and Whole Slide Image segmentation, leveraging the comprehensive set of classes which includes all major morphologies per organ. It serves as a robust benchmark and starting point for further developments in digital pathology research.
    
    \item We open-source both the \textbf{dataset and the baseline models} to the research community, facilitating collaboration, reproducibility, and the advancement of methodologies in digital pathology. 
\end{itemize}

\section{Related Work}

Several patch-level histopathology image datasets have been introduced for classification tasks, each with different annotation types, scales, and focuses. We review the most relevant datasets and compare their characteristics to the proposed SPIDER dataset.

\subsection{Lymph Node Metastasis (Camelyon16/PatchCamelyon)}

Camelyon16 \citep{10.1001/jama.2017.14585camelyon} was one of the earliest large WSI datasets, containing 400 lymph node WSIs for detecting breast cancer metastases. Patches extracted from Camelyon16 form the PatchCamelyon (PCam) dataset \citep{veeling2018rotationequivariantcnnsdigitalpcam}, which includes 327,680 small image patches (96×96 pixels) labeled binary for tumor vs. normal. These labels are derived from expert-annotated tumor regions in the WSIs. 

PCam’s strengths are its large scale and simplicity - it provides a massive training set for binary classification and is easily trainable on a single GPU. It is openly available under a permissive license. However, PCam is limited to a single binary task (metastasis detection in lymph nodes) with no multi-class differentiation.

\subsection{Breast Cancer Histopathology (BreaKHis and BACH)}

Two prominent patch-level datasets target breast biopsy pathology:

\begin{itemize}
    \item \textbf{BreaKHis \citep{7312934breakhis}:} Contains 7,909 microscopic patch images from 82 patients, split into benign vs. malignant tumors (binary) with further 8 subclass labels for specific tumor subtypes. Images in BreaKHis come at four magnifications (40X, 100X, 200X, 400X), each image being approximately 700×460 pixels. All images were annotated by pathologists, ensuring expert label quality.
    \item \textbf{BACH (ICIAR 2018) \citep{ARESTA2019122BACH}:} Contains 400 high-resolution patches (2048×1536 pixels at 20X magnification) categorized into four classes: normal, benign, \textit{in situ} carcinoma, and invasive carcinoma. Each image’s ground truth was confirmed by two expert pathologists, ensuring high annotation quality.
\end{itemize}

BreaKHis’s strengths lie in its inclusion of multiple tumor subtypes and expert-validated labels, enabling both binary and multi-class evaluation. Its multi-magnification nature tests model robustness across scales. However, BreaKHis is relatively small (only a few thousand patches) and focuses only on breast tissue, limiting its diversity. Similarly, BACH images are large and detailed, but the dataset is very small (N=400), making deep learning training challenging without augmentation.
Both BreaKHis and BACH are publicly available under open licenses.

\subsection{Colorectal Tissue Classification (NCT-CRC-HE-100K)}

The NCT-CRC-HE-100K (Kather) \citep{kather_2018_1214456} dataset focuses on colorectal cancer tissue. It consists of 100,000 H\&E-stained image patches (224×224 pixels) extracted from 86 WSIs, along with an additional 7,180 patches from 50 WSIs as an independent validation set. All patches are labeled into nine tissue categories - including tumor epithelium, stroma, immune cells, and various normal tissues - based on manual region delineations by pathologists. The images were color-normalized for consistency.

This dataset’s strengths include its large size (100k), multi-class granularity (covering tumor and microenvironment classes), and high-quality annotations (expert-drawn regions for each class). It is open access (with a Creative Commons license) for easy availability. However, NCT-CRC-HE-100K is focused on a single organ (colon).

\subsection{Multi-Organ Cancer Datasets (LC25000 and Others)}

Multi-organ histopathology datasets are still rare. A notable example is LC25000 \citep{borkowski2019lungcoloncancerhistopathological}, which contains 25,000 images (768×768 pixels) equally divided into five classes: lung adenocarcinoma, lung squamous cell carcinoma, benign lung tissue, colon adenocarcinoma, and benign colon tissue. Each class has 5,000 patches, and all images were validated and made HIPAA-compliant for public release.

LC25000’s strength is its coverage of two organs (lung and colon) and both cancerous and normal tissues, offering a simple multi-class classification task with a balanced dataset. It is freely available (e.g., on Kaggle) for researchers. LC25000’s images are relatively large but fewer in number compared to other datasets, and the number of classes is also limited.

\subsection{Significance of the SPIDER Dataset}

Across existing patch-level datasets, common limitations include:
\begin{itemize}
    \item Narrow scope, often only a single organ
    \item Limited class coverage
    \item Modest dataset size
\end{itemize}

The proposed SPIDER dataset overcomes these gaps. It provides diverse multi-class annotations verified by expert pathologists, ensuring high annotation quality. It is large-scale, on the order of hundreds of thousands of patches, making it larger than the biggest current datasets in sheer size. Importantly, SPIDER covers multiple organ types within one unified dataset, and has a large class coverage per each which is unprecedented among publicly available patch-level histopathology collections. Moreover, since SPIDER has been annotated on a private dataset, it has not been included in the training of existing models, making it particularly valuable for independent validation and benchmarking of such models.

Finally, SPIDER is released under a permissive open license, ensuring broad accessibility for both academic and clinical AI research. In summary, SPIDER’s combination of annotation richness, scale, organ diversity, and availability makes it a significant contribution. It enables more comprehensive training and evaluation of pathology classification models, addressing the weaknesses of earlier datasets and pushing the field toward more generalizable and robust histopathology AI systems.

\section{Dataset}

\subsection{Dataset description}

SPIDER is a brand-new patch-level dataset curated from proprietary whole slide images (WSIs) with expert pathologist annotations. 
Each data point includes a central 224×224 patch captured at 20X magnification, along with a class label. Additionally, it comes with 24 surrounding context patches of the same size, together forming a composite 1120×1120 region.
This context is important because some patches are difficult or even impossible to classify correctly on their own. A simple example is distinguishing between fat tissue and empty background, which is often impossible from 224×224 patch itself. 
And there are many more intricate cases where the context around the patch is required to correctly classify the patch itself. The importance of this context is further explored in Subsection~\ref{subsec:ablation}.

The dataset currently includes four organ types: Skin, Colorectal, Thorax, and Breast. We provide a train-test split to ensure consistent benchmarking, but users can also merge and re-split the data as needed. The split is done at the slide level, meaning patches from the same WSI do not appear in both the training and test sets. We aim for an 80:20 train-test ratio, though the exact numbers vary slightly across tissue classes.

\begin{table}[htb]
\centering
\caption{Dataset composition across organ types. Each central patch is extracted from a WSI and is accompanied by 24 context patches, forming a larger 1120×1120 region. Due to overlaps in context patches, the number of total unique patches is lower than a basic estimate of total central patches × 25.}
\label{tab:dataset-train-val-test-dist}
\begin{tabular}{lrrrrrr}
\toprule
     Organ &  Train & Test & Total Central Patches & Total Unique Patches & Total Slides & Total Classes \\
\midrule
      Skin &   131,164 &    28,690 & 159,854 & 2,696,987 & 3,784 & 24\\
      Colorectal &   63,989 &    13,193 & 77,182 & 1,039,150 & 1,719 & 14 \\
      Thorax &   63,319 &    14,988 & 78,307 & 599,459 & 411 & 14 \\
      Breast &   80,858 &    12,034 & 92,892 & 984,924 & 921 & 18 \\
\bottomrule
\end{tabular}
\end{table}

Table~\ref{tab:dataset-train-val-test-dist} presents the total number of patches for each organ and their train-test split. It also includes the number of WSIs from which the patches were extracted and the total number of classes per organ. A detailed breakdown of tissue classes and sample counts is available in Appendix~\ref{appendix:class_dist_appendix}.

\begin{figure}[htb]
    \centering
    \includegraphics[width=1\linewidth]{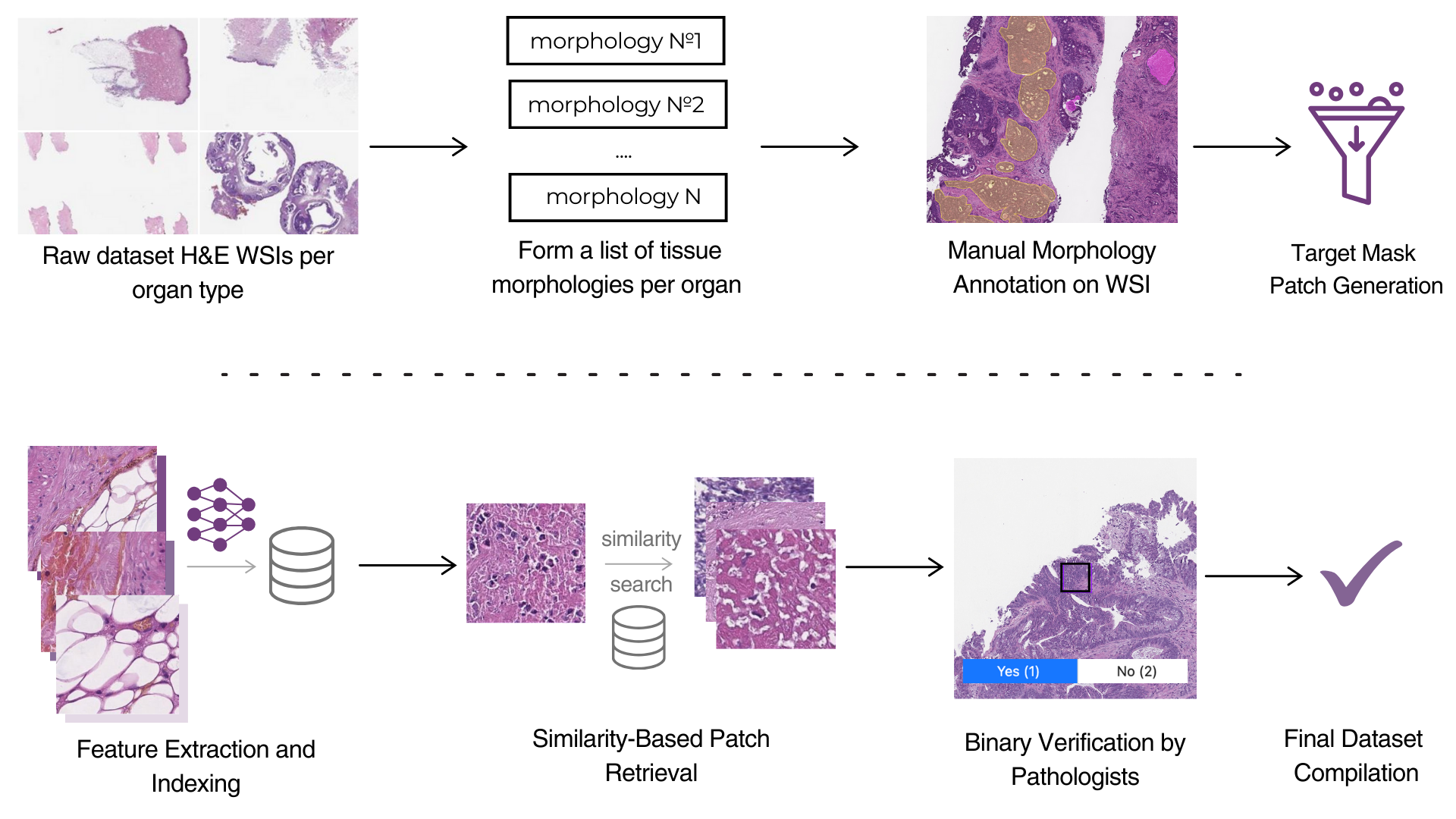}
    \caption{Dataset preparation pipeline: Raw whole-slide images (WSIs) undergo expert annotation, patch extraction, feature embedding, and similarity-based retrieval. A final verification step ensures high-quality labeled patches for training.}
    \label{fig:dataset-prep}
\end{figure}

\subsection{Data Preparation}
\label{subsec:data_preparation}

In order to create a substantial patch-level dataset of high-quality, manually annotated image patches out of the original unannotated dataset we engage professional pathologists to perform the annotations. The process of creating a training dataset for a specific organ type involves the following steps:

\begin{itemize}
    \item \textbf{Raw Dataset Creation:} 
    We compile a raw dataset consisting of H\&E stained WSIs of the same organ type. Then with professional pathologists we form a list of tissue morphologies relevant to that organ, so that it would cover the key morphologies which exist in that organ.

    \item \textbf{Slide Selection and Initial Annotation:}
    WSIs are selected from the raw dataset based on their diagnosis. Pathologists then use the web-based HistAI CELLDX platform to identify and annotate (draw) regions containing the target morphology within the selected slides. Annotations are created using the polygon or brush tools, which enables pathologists to delineate precise mask regions around the region of interest.

    \item \textbf{Target Mask Patch Generation:}
    Using manual annotations from the initial annotation step, we generate non-overlapping 224×224 patches at 20X magnification. Approximately 500–1,000 patches per class are obtained at this stage, serving as target patches for the similarity-based patch retrieval step.

    \item \textbf{Patch Extraction and Feature Indexing:}
    Each WSI in the raw dataset is divided into 224×224 pixel non-overlapping patches. White background regions are excluded using Otsu thresholding. Each patch is then processed with the Hibou-L model to extract feature embeddings, which are stored in a Faiss index~\citep{douze2024faiss} for efficient similarity-based searches.

    \item \textbf{Similarity-Based Patch Retrieval:}
    Utilizing the Faiss index, we perform similarity-based searches to identify patches across slides from raw dataset that closely resemble the target patches. This approach efficiently expands our dataset by identifying visually comparable patches from a diverse set of slides. The retrieved patches are then filtered by pathologists.

    \item \textbf{Binary Verification by Pathologists:}
    To ensure data quality, pathologists conduct a binary labeling task using HistAI patch annotation platform. Each patch is presented within a 2016×2016 pixel context to provide spatial information. Pathologists determine whether the central patch belongs to the target class (e.g., Nevus), enabling them to verify thousands of patches per hour with high accuracy.

    \item \textbf{Final Dataset Compilation:}
    Upon completing the verification process, a sufficiently large curated dataset of annotated patches is obtained. This dataset is then used to train models.
\end{itemize}

\section{Model}
\label{sec:model}
\subsection{Model architecture}

We design our model architecture with the aim of effectively incorporating context patches surrounding each central patch, improving the classification accuracy of the central patch.

\begin{figure}[htb]
    \centering
    \includegraphics[width=1\linewidth]{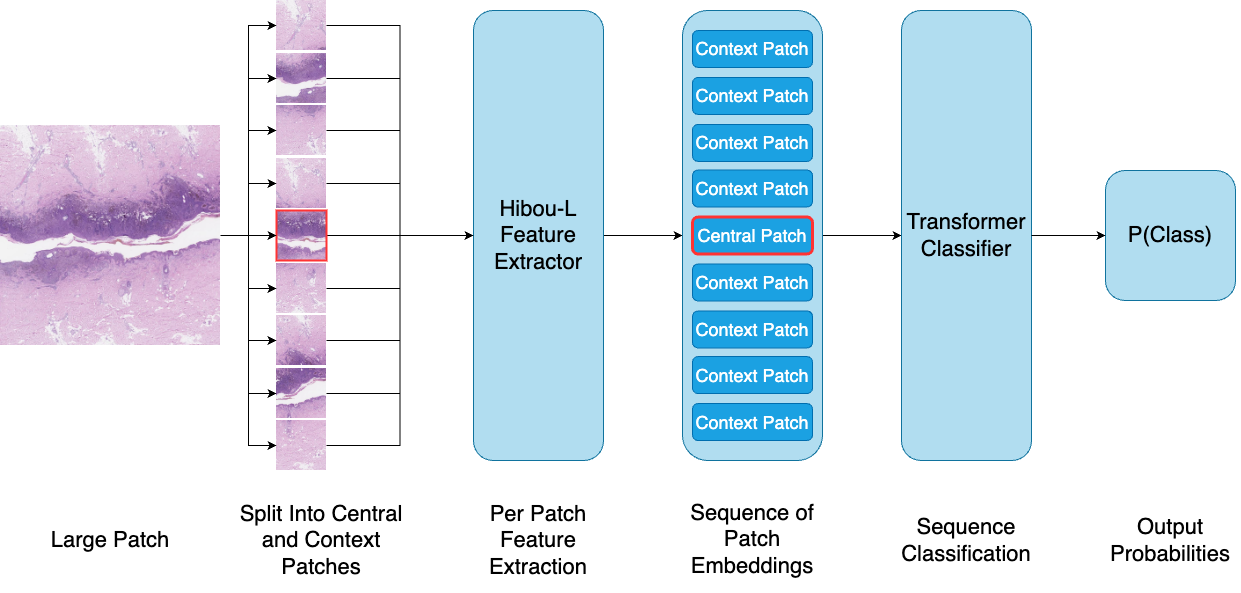}
    \caption{Model architecture overview: The classifier processes a central patch alongside surrounding context patches. Features are extracted using the Hibou-L model, and an attention-based classification head integrates context information to improve central patch classification.}
    \label{fig:classifier}
\end{figure}

The proposed model consists of two main components: Hibou-L vision foundation model \citep{nechaev2024hiboufamilyfoundationalvision} as a feature extractor and an attention-based classification head, as shown in Figure~\ref{fig:classifier}.

The input to the model is a large patch of size 1120 × 1120 pixels, which is divided into a 5 × 5 grid of smaller patches, each measuring 224 × 224 pixels. These smaller patches are processed individually by the feature extractor, generating embeddings for each patch.

The central patch in the grid is the primary target for classification and the surrounding patches, known as context patches, provide additional information to help the model capture a broader view improving its ability to classify the central patch.
The embeddings from all patches (central and context) are stacked together and passed to the attention-based classification head. The attention mechanism enables the model to incorporate information from context patches for class prediction of the central patch.

For each organ type, we train a separate model on multi-class classification task. The model is optimized using a cross-entropy loss function, which is a standard method for handling multiple class labels. During training, we keep the Hibou feature extractor frozen and only train the classification head. This ensures that the model relies on the high-quality features learned during pretraining rather than modifying them which reduces overfitting to specific images.

\subsection{Results}

The trained models demonstrate strong performance in the multi-class classification task. Table~\ref{tab:metrics_main} presents the overall accuracy, precision, and F1 score for each organ on the test set. Detailed per-class metrics can be found in Appendix~\ref{appendix:det_metrics}. A detailed description of the training configuration, including hyperparameter settings is provided in Appendix~\ref{appendix:training_details}

\begin{table}[htb]
\centering
\caption{Performance metrics of models across different organs on the test set. Accuracy, Precision, and F1 score are reported.}
\label{tab:metrics_main}
\begin{tabular}{lrrr}
\toprule
     Organ &  Accuracy &  Precision & F1 \\
\midrule
      Skin &   0.940 &    0.935    & 0.937 \\
      Colorectal &   0.914 & 0.917 & 0.915 \\
      Thorax &   0.962 &         0.958 & 0.960 \\
      Breast & 0.902 & 0.896 & 0.897 \\
\bottomrule
\end{tabular}
\end{table}

\subsection{Ablation Study}
\label{subsec:ablation}

Context patches play a crucial role in enhancing model performance. To evaluate impact of the context, we trained models with different context sizes and assessed their accuracy on the test set. As shown in Table~\ref{tab:metrics_ablation}, reducing the context size leads to a decline in model accuracy across all organs. The decrease is more pronounced when context is completely removed, demonstrating the importance of spatial context in model predictions.

\begin{table}[htb]
\centering
\caption{Impact of context size on model accuracy across different organs. Larger context windows improve accuracy, emphasizing the importance of contextual information.}
\label{tab:metrics_ablation}
\begin{tabular}{lccc}
\toprule
     Organ &  Large Context (1120×1120) &  Medium Context (672×672) &  No Context (224×224) \\
\midrule
      Skin       &  0.940 &  0.935 &  0.923 \\
      Colorectal &  0.914 &  0.906 &  0.895 \\
      Thorax     &  0.962 &  0.960 &  0.956 \\
      Breast     &  0.902 &  0.895 &  0.883 \\
\bottomrule
\end{tabular}
\end{table}

\subsection{Usecases}

The model was trained on a classification task, yet due to comprehensive class coverage in SPIDER, a trained model's applications go beyond simple patch classification:
\begin{itemize}
    \item \textbf{Rapid Identification and Explainable AI for Pathologists.}
    The patch-level classification predictions can be aggregated into a coarse segmentation mask, enabling pathologists to quickly locate potentially malignant or otherwise significant areas in a Whole Slide Image (WSI). This visualization serves as a heat map overlay, providing an intuitive representation of the model’s predictions. By highlighting abnormal regions, the model speeds up the review process, reduces the time to diagnosis, and enhances interpretability. Clinicians and researchers can validate AI-driven suggestions more efficiently, improving diagnostic confidence and workflow efficiency.
    Examples of a WSI segmentation can be found at Appendix~\ref{appendix:full-slide-seg}.

    \item \textbf{Quantitative Metrics for Research and Treatment Planning.}  
    Beyond simple detection, the model’s patch-level insights enable automatic calculation of proportions such as tumor area, stromal content, and other morphological components. These metrics can guide both clinical decisions (e.g., selecting patients for targeted therapies) and research studies (e.g., identifying biomarkers correlated with treatment response).

    \item \textbf{Foundation for Multimodal and More Advanced Models.}  
    A slide-level mosaic of patch labels yields a highly detailed representation of tissue morphology. This labeled output can be used to train or augment multimodal models, including vision-language systems that require large amounts of paired text-image data. By automatically generating such pairs, the approach scales the development of richer AI solutions.
\end{itemize}

\section{Discussion}

Our work introduces SPIDER, a comprehensively annotated multi-organ histopathology dataset that addresses significant limitations in the computational pathology field. Through our experimentation and model development, we have identified several key insights and implications for future research.

\begin{itemize}
    \item \textbf{The Value of Supervised Approaches with Foundation Models} \\
    Our supervised approach, leveraging foundation models, demonstrates strong performance. By creating an expert-annotated patch-level dataset and fine-tuning a pretrained foundation model, we achieved high classification accuracy across multiple tissue types without requiring extremely large datasets. This approach effectively utilizes the generalization capabilities of foundation models.
    
    \item \textbf{The Critical Role of Contextual Information} \\
    The ablation study demonstrates that classification accuracy progressively decreases as contextual information is reduced, with the most pronounced effect occurring when context is completely removed. 
    This finding aligns with how pathologists operate in clinical practice; they rarely examine tissue in isolation but consider surrounding structures to make accurate assessments. By incorporating a larger 1120×1120 context window, our model can better mimic the holistic assessment approach used by expert pathologists.
\end{itemize}

\subsection{Limitations and Future Work}

Despite the promising results, several limitations and opportunities for future work remain. First, while SPIDER covers multiple organ types, expanding to additional organs would further enhance its utility. Second, the current model architecture, while effective, could potentially be improved through more sophisticated attention mechanisms or alternative approaches to incorporating contextual information.
Additionally, exploring the integration of clinical metadata with image features could enhance model performance and provide more clinically relevant insights.

Future work should also investigate the model's generalizability across different scanning systems, staining protocols, and patient populations. While our models demonstrate strong performance on the test set, broader validation across diverse real-world settings would further establish their robustness and clinical utility.

\subsection{Conclusion}

SPIDER represents a significant contribution to the field of computational pathology, providing a comprehensive, multi-organ dataset with expert-validated annotations and crucial contextual information. Our baseline models demonstrate the effectiveness of combining foundation models with supervised learning approaches and highlight the importance of spatial context in accurate tissue classification.

By making both the dataset and models publicly available, we aim to accelerate research in digital pathology and enable the development of more robust, clinically relevant AI tools. The promising results across diverse tissue types suggest that this approach could be extended to other organ systems, potentially transforming pathology workflows and enhancing both diagnostic accuracy and efficiency.

\section*{Acknowledgements}

This work would not have been possible without the dedication and expertise of our incredible team of pathologists:  
Elena Konovalova, Daria Shushkanova, Antonina Vedinova, Vasilii Fedotov, Galina Makarniaeva, and Aliaksandr Zhurauliou.  

Your meticulous annotations, patience, and deep knowledge brought SPIDER to life. We are truly grateful for the countless hours you spent ensuring the accuracy and quality of this dataset. Thank you for your hard work, attention to detail, and for making this project a reality.

\bibliographystyle{unsrtnat}
\bibliography{references}

\newpage
\appendix
\section{Appendix}
\subsection{Class distribution in organs}
\label{appendix:class_dist_appendix}
\renewcommand{\thefigure}{\thesection\arabic{figure}} 
\renewcommand{\thetable}{\thesection\arabic{table}} 
\setcounter{figure}{0} 
\setcounter{table}{0}

\begin{figure}[htb]
    \centering
    \includegraphics[width=1\linewidth]{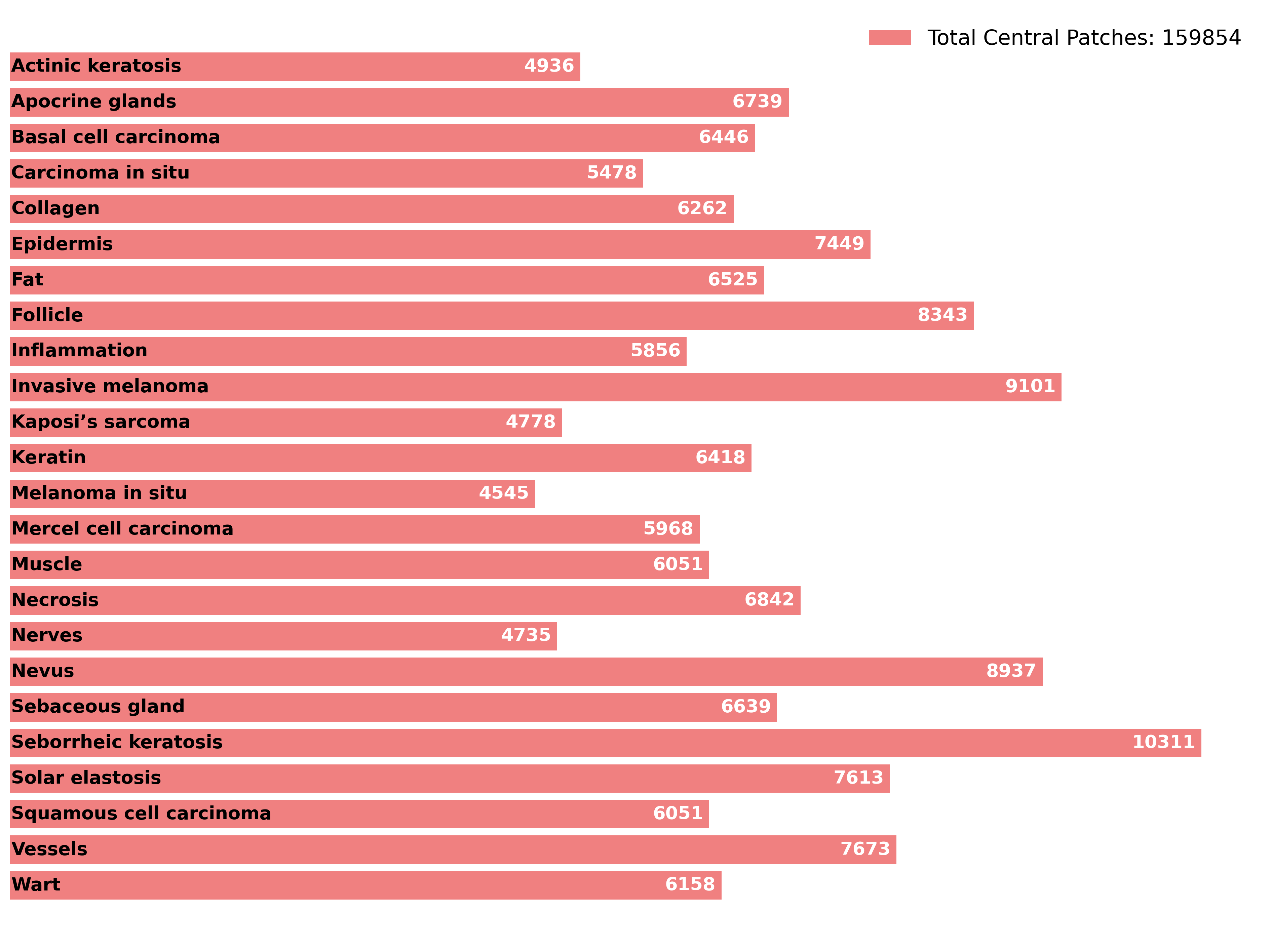}
    \caption{Dataset skin class distribution}
    \label{fig:skin-distribution}
\end{figure}

\begin{figure}[htb]
    \centering
    \includegraphics[width=1\linewidth]{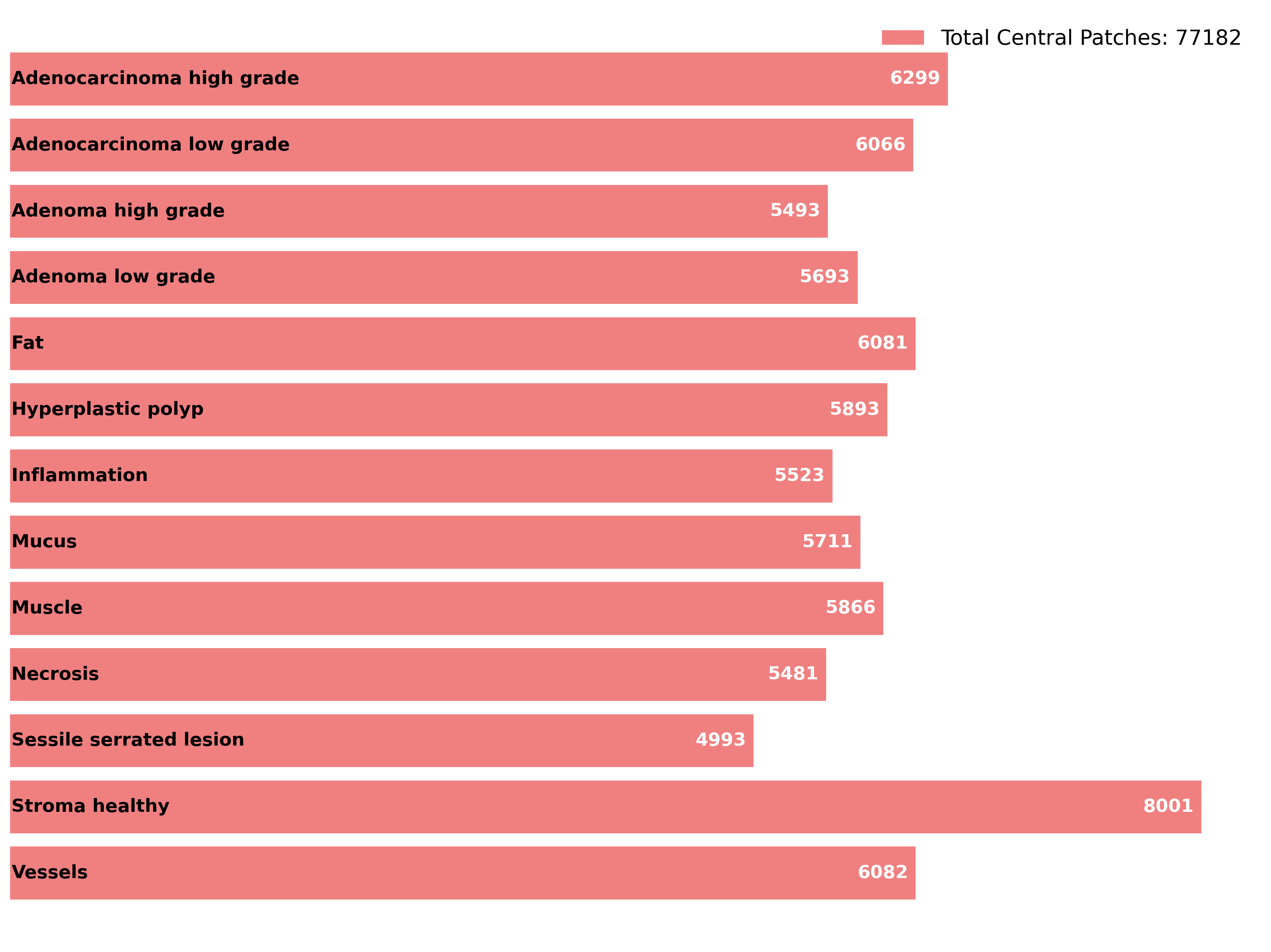}
    \caption{Dataset colorectal class distribution}
    \label{fig:colorectal-distribution}
\end{figure}

\begin{figure}[htb]
    \centering
    \includegraphics[width=1\linewidth]{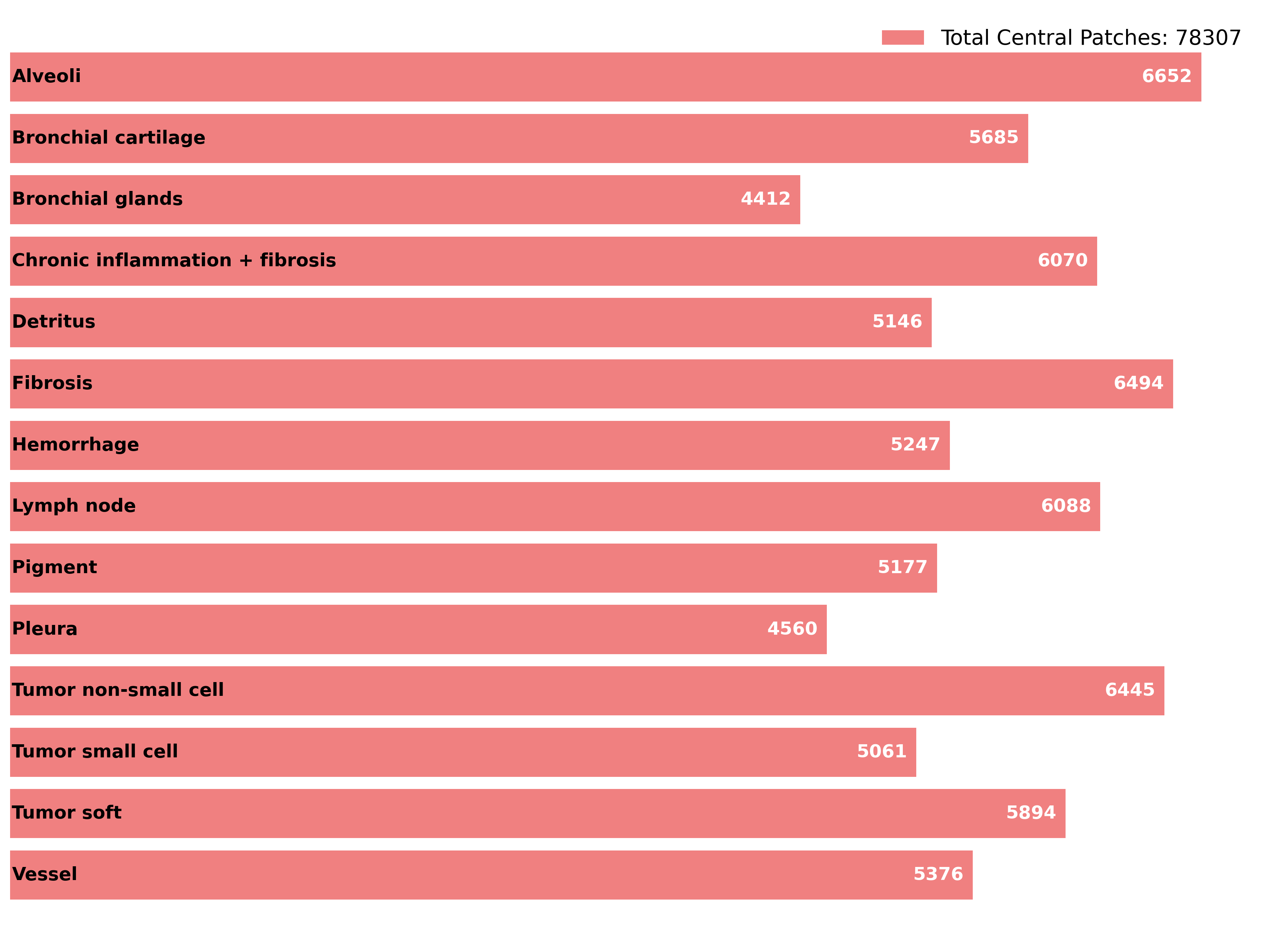}
    \caption{Dataset thorax class distribution}
    \label{fig:thorax-distribution}
\end{figure}

\begin{figure}[htb]
    \centering
    \includegraphics[width=1\linewidth]{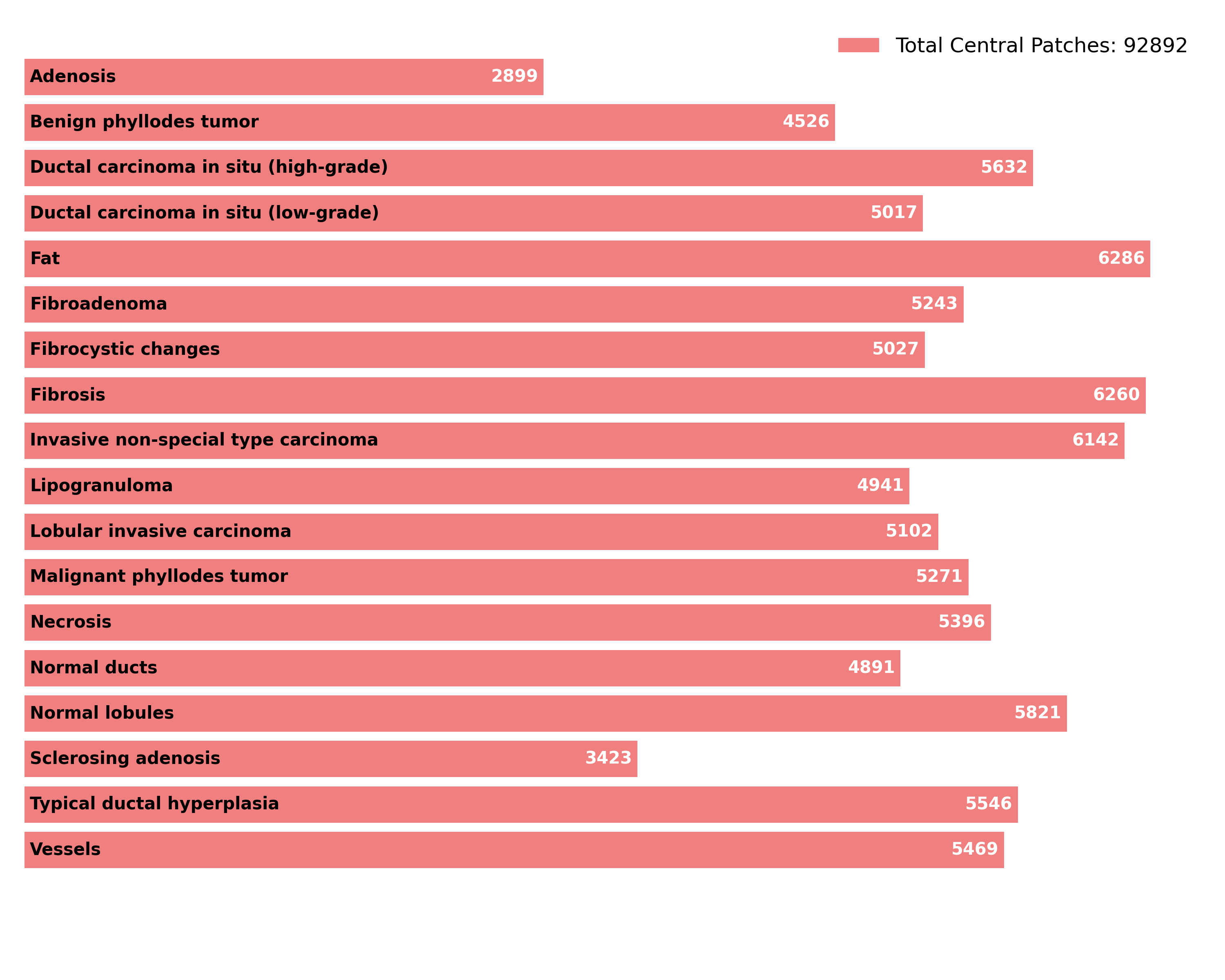}
    \caption{Dataset breast class distribution}
    \label{fig:breast-distribution}
\end{figure}

\FloatBarrier

\subsection{Training details}
\label{appendix:training_details}

\begin{table}[h]
    \centering
    \caption{Training hyperparameters}
    \label{table:training_params}
    \begin{tabular}{l l}
        \toprule
        \textbf{Parameter} & \textbf{Value} \\
        \midrule
        Epochs & 10 \\
        Batch size & 256 \\
        Loss function & Cross entropy \\
        Label smoothing & 0.2 \\
        Optimizer & AdamW~\citep{loshchilov2019decoupledadamw} \\
        Learning rate & \(4 \times 10^{-4}\) \\
        Weight decay & 0.01 \\
        Learning rate scheduler & Linear warmup + Cosine annealing \\
        Warmup epochs & 1 \\
        Mixed precision & FP16 \\
        \bottomrule
    \end{tabular}
\end{table}

\begin{table}[h]
    \centering
    \caption{Model configuration}
    \label{table:model_config}
    \begin{tabular}{l l}
        \toprule
        \textbf{Parameter} & \textbf{Value} \\
        \midrule
        Feature extractor & Hibou-L \\
        Classification head & Bert~\citep{devlin2019bertpretrainingdeepbidirectional} \\
        Hidden size & 128 \\
        Number of hidden layers & 1 \\
        Number of attention heads & 1 \\
        Intermediate size & 128 \\
        Maximum position embeddings & 25 \\
        Hidden dropout probability & 0.5 \\
        Attention dropout probability & 0.3 \\
        Head dropout probability & 0.3 \\
        \bottomrule
    \end{tabular}
\end{table}

\FloatBarrier
\newpage
\subsection{Detailed metrics}
\label{appendix:det_metrics}

\begin{table}[htb]
\centering
\begin{tabular}{lccc}
\hline
\textbf{Class} & \textbf{Accuracy} & \textbf{Precision} & \textbf{F1} \\
\hline
Actinic Keratosis & 0.768 & 0.817 & 0.792 \\
Apocrine Glands & 0.999 & 0.999 & 0.999 \\
Basal Cell Carcinoma & 0.959 & 0.913 & 0.935 \\
Carcinoma In Situ & 0.761 & 0.698 & 0.728 \\
Collagen & 0.989 & 0.992 & 0.990 \\
Epidermis & 0.871 & 0.933 & 0.901 \\
Fat & 0.997 & 0.998 & 0.997 \\
Follicle & 0.942 & 0.953 & 0.947 \\
Inflammation & 0.926 & 0.974 & 0.949 \\
Invasive Melanoma & 0.936 & 0.937 & 0.937 \\
Kaposi’s Sarcoma & 0.990 & 0.906 & 0.946 \\
Keratin & 0.994 & 0.977 & 0.985 \\
Melanoma In Situ & 0.976 & 0.962 & 0.969 \\
Mercel Cell Carcinoma & 0.887 & 0.998 & 0.939 \\
Muscle & 0.984 & 0.984 & 0.984 \\
Necrosis & 0.981 & 0.954 & 0.967 \\
Nerves & 0.999 & 1.000 & 0.999 \\
Nevus & 0.973 & 0.981 & 0.977 \\
Sebaceous Gland & 0.987 & 0.984 & 0.985 \\
Seborrheic Keratosis & 0.929 & 0.914 & 0.922 \\
Solar Elastosis & 0.997 & 0.988 & 0.993 \\
Squamous Cell Carcinoma & 0.839 & 0.826 & 0.832 \\
Vessels & 0.991 & 0.991 & 0.991 \\
Wart & 0.881 & 0.772 & 0.823 \\
\midrule
Total & 0.940 & 0.935 & 0.937 \\
\hline
\end{tabular}
\caption{Extended classification metrics for skin model.}
\label{tab:classification_metrics_skin}
\end{table}

\begin{table}[htb]
\centering
\begin{tabular}{lccc}
\hline
\textbf{Class} & \textbf{Accuracy} & \textbf{Precision} & \textbf{F1} \\
\hline
Adenocarcinoma High Grade & 0.861 & 0.963 & 0.909 \\
Adenocarcinoma Low Grade & 0.819 & 0.848 & 0.833 \\
Adenoma High Grade & 0.805 & 0.762 & 0.783 \\
Adenoma Low Grade & 0.915 & 0.865 & 0.889 \\
Fat & 0.994 & 0.997 & 0.995 \\
Hyperplastic Polyp & 0.833 & 0.866 & 0.850 \\
Inflammation & 0.978 & 0.959 & 0.969 \\
Mucus & 0.895 & 0.818 & 0.855 \\
Muscle & 0.981 & 0.970 & 0.976 \\
Necrosis & 0.977 & 0.976 & 0.977 \\
Sessile Serrated Lesion & 0.889 & 0.961 & 0.924 \\
Stroma Healthy & 0.977 & 0.970 & 0.974 \\
Vessels & 0.961 & 0.969 & 0.965 \\
\midrule
Total & 0.914 & 0.917 & 0.915 \\
\hline
\end{tabular}
\caption{Extended classification metrics for colorectal model.}
\label{tab:classification_metrics_colorectal}
\end{table}

\begin{table}[htb]
\centering
\begin{tabular}{lccc}
\hline
\textbf{Class} & \textbf{Accuracy} & \textbf{Precision} & \textbf{F1} \\
\hline
Alveoli & 0.986 & 0.926 & 0.955 \\
Bronchial Cartilage & 1.000 & 1.000 & 1.000 \\
Bronchial Glands & 0.995 & 1.000 & 0.998 \\
Chronic Inflammation + Fibrosis & 0.950 & 0.998 & 0.973 \\
Detritus & 0.961 & 0.959 & 0.960 \\
Fibrosis & 0.932 & 0.918 & 0.925 \\
Hemorrhage & 0.948 & 0.988 & 0.968 \\
Lymph Node & 0.962 & 0.994 & 0.978 \\
Pigment & 0.935 & 0.863 & 0.898 \\
Pleura & 0.914 & 0.892 & 0.903 \\
Tumor Non-Small Cell & 0.995 & 0.997 & 0.996 \\
Tumor Small Cell & 1.000 & 0.993 & 0.996 \\
Tumor Soft & 1.000 & 1.000 & 1.000 \\
Vessel & 0.887 & 0.885 & 0.886 \\
\midrule
Total & 0.962 & 0.958 & 0.960 \\
\hline
\end{tabular}
\caption{Extended classification metrics for thorax model.}
\label{tab:classification_metrics_thorax}
\end{table}

\begin{table}[htb]
\centering
\begin{tabular}{lccc}
\hline
\textbf{Class} & \textbf{Accuracy} & \textbf{Precision} & \textbf{F1} \\
\hline
Adenosis & 0.722 & 0.664 & 0.691 \\
Benign Phyllodes Tumor & 0.967 & 0.998 & 0.983 \\
Ductal Carcinoma In Situ (High-Grade) & 0.993 & 0.977 & 0.985 \\
Ductal Carcinoma In Situ (Low-Grade) & 0.905 & 0.887 & 0.896 \\
Fat & 0.999 & 0.999 & 0.999 \\
Fibroadenoma & 0.987 & 0.973 & 0.980 \\
Fibrocystic Changes & 0.769 & 0.576 & 0.659 \\
Fibrosis & 0.998 & 0.989 & 0.994 \\
Invasive Non-Special Type Carcinoma & 0.988 & 1.000 & 0.994 \\
Lipogranuloma & 0.997 & 0.991 & 0.994 \\
Lobular Invasive Carcinoma & 0.987 & 0.980 & 0.984 \\
Malignant Phyllodes Tumor & 1.000 & 1.000 & 1.000 \\
Necrosis & 0.984 & 0.999 & 0.991 \\
Normal Ducts & 0.902 & 0.817 & 0.858 \\
Normal Lobules & 0.660 & 0.625 & 0.642 \\
Sclerosing Adenosis & 0.727 & 0.778 & 0.752 \\
Test & 0.902 & 0.896 & 0.897 \\
Typical Ductal Hyperplasia & 0.679 & 0.882 & 0.767 \\
Vessels & 0.971 & 0.989 & 0.980 \\
\midrule
Total & 0.902 & 0.896 & 0.897 \\
\hline
\end{tabular}
\caption{Extended classification metrics for breast model.}
\label{tab:classification_metrics_breast}
\end{table}

\FloatBarrier

\subsection{Full-slide segmentation examples}
\label{appendix:full-slide-seg}
\begin{figure}[htb]
    \centering
    \includegraphics[width=1\linewidth]{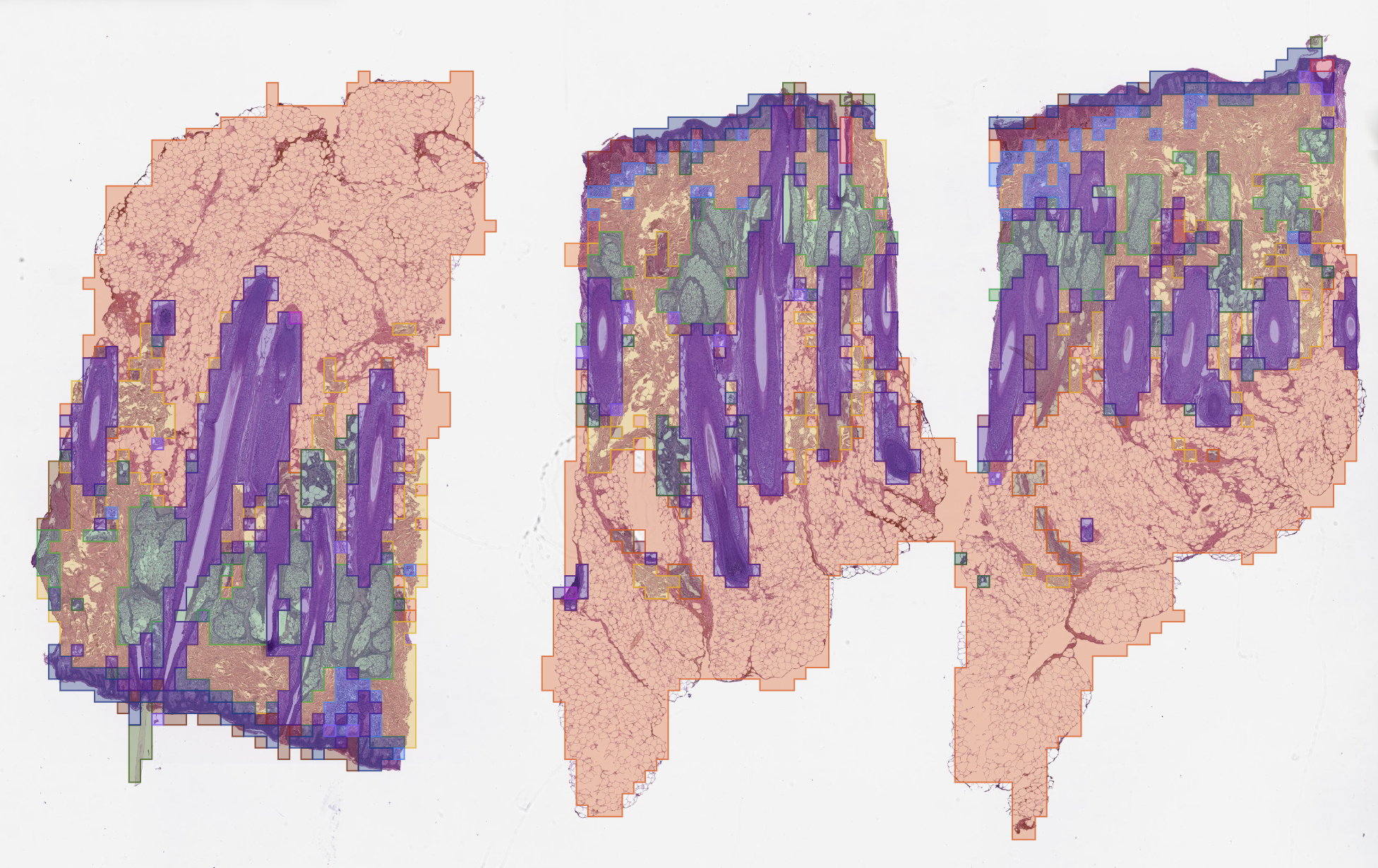}
    \caption{Example of a full slide segmentation. Each color represents a separate class.}
    \label{fig:full-slide-seg1}
\end{figure}


\end{document}